\documentstyle[12pt]{article}

\newcommand{\eq}[1]{(\ref{#1})}
\newcommand{\beq}{\begin{equation}}
\newcommand{\eeq}{\end{equation}}
\newcommand{\barr}{\begin{array}}
\newcommand{\earr}{\end{array}}
\newcommand{\beqa}{\begin{eqnarray}}
\newcommand{\eeqa}{\end{eqnarray}}

\newcommand{\nn}{\nonumber}
\newcommand{\ra}{\rightarrow}
\newcommand{\ket}[1]{\vert\,{#1}\,\rangle}

\newcommand{\pl}{\partial}
\newcommand{\Uq}{{\cal U}_q(sl(2))}
\newcommand{\gz}{ 1+z\bar{z} }
\newcommand{\zb}{ \bar{z}}
\newcommand{\zt}{\tilde{z}}

\begin{document}
\topmargin 0pt
\oddsidemargin 1mm
\begin{titlepage}
\setcounter{page}{0}
\begin{flushright}
LMU-TPW 94-19\\
OWUAM-005\\
hep-th 9412017\\
\end{flushright}
\vspace{15mm}
\begin{center}
{\Large{\bf Laughlin States on the Sphere as Representations of $\Uq$}}

\vspace{2cm}
{\large Naruhiko Aizawa
\footnote{e-mail: aizawa@miho.rcnp.osaka-u.ac.jp}}\\
{\em Department of Applied Mathematics, Osaka Women's University\\
Sakai, Osaka 590, Japan}\\
and\\
{\large Sebastian Sachse,
\footnote{e-mail: sachse@lswes8.ls-wess.physik.uni-muenchen.de}
Haru-Tada Sato
\footnote{JSPS Fellow, on leave of absence from
Dept. Physics, Osaka Univ.\\
\phantom{mil}e-mail: hsato@lswes8.ls-wess.physik.uni-muenchen.de}}\\
{\em Sektion Physik der Universit\"at M\"unchen \\
 Lehrstuhl Professor Wess, Theresienstra\ss e 37\\
 D-80333 M\"unchen, Germany }
\end{center}

\vspace{7mm}

\begin{abstract}
We discuss quantum algebraic structures of the systems of electrons or
quasiparticles on a sphere of which center a magnetic monople is located
on. We verify that the deformation parameter is related to the filling
ratio of the particles in each case.
\end{abstract}
\vspace{1cm}
\end{titlepage}
\newpage
\renewcommand{\thefootnote}{\arabic{footnote}}
\indent

The charged particle systems confined on a two-dimensional surface
in a strong magnetic field show the fractional quantum Hall effect
\cite{QHE}. The maximal symmetry of the systems is the group of magnetic
translations and making use of the generators, we can compose a quantum algebra
$\Uq$ \cite{Fad}. When the deformation parameter $q$ is a real number,
the representations of $\Uq$ is essentially equivalent to those of the
angular momentum algebra, which corresponds to a special limit of $\Uq$,
i.e., $q\ra1$. In this case, $q$ is nothing more than an artificial
parameter, however we have no reason to choose $q=1$ apriori. We hence
inquire the meaning of the deformation parameter.

One possible interpretation is to incorporate a physical quantity into
the deformation parameter and we can remove the freedom to choose the
value of $q$ accordingly. This situation becomes possible when we consider
the case of $q$ a complex number. Namely, it may be represented as
\beq
                 q=\exp(2\pi i \nu_e),         \label{eq1}
\eeq
where $\nu_e$ is the filling factor for electrons. This relation is verified
in the planar (cylinder) and torus cases \cite{HS}. The quantum group symmetry
of these cases can be discussed in common manners using magnetic
translations on a square lattice. On the other hand,
how is the case of spherical \cite{sphere,nu,Li} quantum Hall systems ?
The realization of $\Uq$ symmetry on a sphere may show a different
feature from previous cases. The maximal symmetry on a sphere is a
rotational one and it does not permit us to construct $\Uq$ in the same
way as the former cases. We therefore discuss a $\Uq$ symmetry on a sphere
and would like to verify the relation \eq{eq1}.

In this paper, we first show the $\Uq$ symmetry with the relation \eq{eq1}
in electrons' wavefunction bases. We start the discussion from
one-particle system in order to clarify how different from others
the spheric case is. After generalizing the argument into many-particles'
system, we discuss the case of quasiparticles.

Let us consider a sphere (the radius $R=1/2$) with the Dirac monopole
located at its center giving rise to a total magnetic flux $\phi$ measured
in units of the flux quantum. Making use of the stereographic projection,
the Hamiltonian of an electron (mass $m_e$) on this sphere reads
\beq
         H ={1\over m_e}(\gz)^2\{D,{\bar D}\}
           ={2\over m_e}(\gz)^2D\bar{D}-{\phi\over m_e},   \label{eq2}
\eeq
where
\beq
D = \pl_z - \frac{\phi}{2} \frac{\zb}{\gz}, \hskip 30pt
{\bar D} = \pl_{\zb} + \frac{\phi}{2} \frac{z}{\gz}.        \label{eq3}
\eeq
Here we should notice that the system \eq{eq2} is very similar
to a supersymmetric Hamiltonian \cite{Ingo}. We have chosen the gauge such
that the Dirac string is at $z=\infty$. In the following argument, we
subtract the constant term $\phi/m_e$ from the above Hamiltonian. The
ground states
\beq
\ket{m} = \psi_m(z,\zb) = \frac{z^m}{(\gz)^{\phi/2}}
                   \hskip 20pt  (0 \leq m \leq \phi)        \label{eq4}
\eeq
are degenerate with all solutions of $\bar{D} \psi = 0 $ in
the Hilbert space endowed with the scalar product
\beq
 <\psi_1,\psi_2>
  = \int \frac{dzd\zb}{(\gz)^2} \bar{\psi_1}{\psi_2}.       \label{eq5}
\eeq
We here omit the orthonormalization factors, which are irrelevant to
our discussions. The degeneracy stems from rotational invariance of the
Hamiltonian so that the angular momentum operator
\beq
             J = z \pl_z - \zb \pl_{\zb}              \label{eq6}
\eeq
can be taken for a complete classification of the ground states by their
magnetic quantum number. With the aid of $J$, we can construct a
representation of $\Uq$ on the states $\ket{m}$ as follows:
\beq
E^+ = - z [J+1]_q, \qquad E^- = z^{-1} [J]_q,
                  \qquad k = q^{J +\frac{1}{2}},      \label{eq7}
\eeq
where
\beq
  [x]_q = \frac{q^x - q^{-x}}{q-q^{-1}}.            \label{eq8}
\eeq
We can easily verify that
\beq
   [ E^+, E^-]\ket{m} =\frac{k^2 - k^{-2}}{q -q^{-1}} \ket{m},
\hskip 30pt
   k E^{\pm}k^{-1}\ket{m} = q^{\pm} E^{\pm}\ket{m}.        \label{eq9}
\eeq
These generators \eq{eq7} are expressed in terms of the operator $J$
which is associated with the rotational symmetry. In the cases of planar
and torus $\Uq$, the generators are written in translational operators
\cite{HS}. Even if taking a mapping from $z$-plane to other coordinates
(for example to a cylinder), \eq{eq7} does not coincide with the
generators of Ref.\cite{HS}. This is the main difference between the
spheric case and others. Another difference is that the generators \eq{eq7}
satisfy the commutation relations
\beq
   [ E^+, E^-] =\frac{k^2 - k^{-2}}{q -q^{-1}},
\hskip 30pt
   k E^{\pm} k^{-1}  = q^{\pm} E^{\pm}         \label{eq10}
\eeq
with the exception of the origin $z=0$ on the projective space. Namely,
the formula $[\pl_{\zb}, \frac{1}{z}] = \delta^2(z)$ does not allow us to
calculate the relations of $\Uq$ directly using \eq{eq7}. Fortunately, we
can safely neglect this kind of singularities when we estimate the
commutation relations on the ground states. In this sense
similarly, we have a Casimir operator
\beq
   C={qk^2+q^{-1}k^{-2} \over (q-q^{-1})^2} +E^-E^+    \label{eq11}
\eeq
which commutes with all the generators and the Hamiltonian;
\beq
  [C,E^\pm]\cong0,\qquad [C,k]\cong0,\qquad [C,H]\cong0,\qquad \label{eq12}
\eeq
where $\cong$ means that the equality holds on the ground states.
We can check also the commutativity btween $\Uq$ and the Hamiltonian
\beq
  [ H,E^\pm]\cong0,\qquad [ H,k]\cong0.   \label{eq13}
\eeq
For fixing $q$, we require that $E^+\ket{\phi}=0$. This condition is satisfied
if we take
\beq
                q = \exp(\frac{2\pi i}{\phi + 1}),       \label{eq14}
\eeq
which is exactly corresponds to \eq{eq1} because $\nu$ is given by
$1/(\phi+1)$ for one particle states on a sphere \cite{nu,Ingo}.

In the next step, we apply the above arguments to $N_e$ electrons on the
sphere. We first mention a free particle picture \cite{Ingo}. Namely,
consider the monopole field to be so strong that we can approximately neglect
electron-electron interactions, in which the Hamiltonian is simply the
sum of $N_e$ single electron Hamiltonians. In this case, the Laughlin
wavefuctions \cite{laugh} become the ground states \cite{Ingo}
\beq
 \Psi_m = \prod_{i < j} (z_i - z_j)^m \prod_i
          ( 1+z_i \bar{z}_i)^{-\phi/2},              \label{eq15}
\eeq
where we omit anti-holomorphic parts and $m$ is an odd interger.
Similarly to one particle case, we have the following restriction from
normalizability condition \cite{Ingo}
\beq
           0 \le m \le \frac{\phi}{N-1}.            \label{eq16}
\eeq
Again $J$ can be used to construct a representation of $\Uq$ on the
ground states $\ket{m}=\Psi_m$ adding bosonic sector (even $m$) in the
multiplet. The generators of $\Uq$ are now in turn
\beqa
E^+&=&- \prod_{i<j}^{N_e} (z_i-z_j) [\tilde{J}+\alpha+\beta]_q, \nn \\
E^-&=&\quad \prod_{i<j}^{N_e}(z_i-z_j)^{-1}
                   [\tilde{J}+\alpha-\beta]_q,\\            \label{eq17}
k &=& q^{\tilde{J} +\alpha}, \nn
\eeqa
where ${\tilde J}$ is defined by the sum of each particle's $J$ as
\beq
 \tilde{J} = \frac{2}{N_e(N_e-1)} \sum_{i=1}^{N_e} J^{(i)}. \label{eq18}
\eeq
The parameter $\alpha$ and $\beta$ are determined later in conformity with
the representation. The $E^+$ and $E^-$ play the role of 'supercharges'
instead of $D$ and ${\bar D}$ as well as $\Uq$ raising-lowering
operators. The deformation parameter is determined similarly as in
the previous case choosing $\alpha=\beta=1/2$
\beq
  q = \exp\left({2\pi i\over 1+{\phi\over N_e-1}}\right).   \label{eq19}
\eeq

Now we can consider the unique Laughlin state which is non-degenerate
(fermionic) single state in the situation of Coulomb's interactions. In
this case, we can impose the rotational invariance on the state. Then the
possible number of values of $m$ reduces to only one defined by \cite{Ingo,Li}
\beq
                   \phi=m(N_e-1).                \label{eq20}
\eeq
In order to see the coincidence with \eq{eq1}, two kind of interpretations
are possible in accordance with the roles of $E^{\pm}$ as whether
'supercharges' or $\Uq$ operators. As for $\Uq$ symmetry,
the Laughlin state is singlet for the $\alpha=\beta=0$ representation
and thus the singlet condition $E^{\pm}\ket{m}=0$ determines the
deformation parameter as
\beq
         q = \exp(2\pi i/m),         \label{eq22}
\eeq
which is in agreement with Laughlin's argument $\nu_e=1/m$.
This result coincides with \eq{eq1}.

As a remark, we make mention of another interpretation which is based on
the analogy of supersymmetry (although we do not have any supersymmetry).
If we follow in the above step (representation $\alpha=\beta=1/2$), we
suppose $E^{\pm}$ not as quantum group generators but as 'supercharges'.
Let us consider to decouple the bosonic sector $\ket{m-1}$ from the doublet
\beq
          E^+\ket{m-1}=E^-\ket{m}=0.              \label{eq21}
\eeq
These conditions mean that each sector can not be transformed by the
'supercharges' $E^+$ and $E^-$ into each other. Imposing these conditions,
we exactly obtain \eq{eq22}. To intuitively obtain the result \eq{eq22}
is to substitute \eq{eq20} into \eq{eq19} omitting the $1$ in the
denominator. The removal of the unity corresponds to exclude the
bosonic sector.

Finally, we discuss that the microscopic $N_s$ quasiparticle (or -hole)
wavefunctions \cite{Halp}, called pseudo wavefunctions, on the sphere
exhibit an $\Uq$ symmetry with similar relation to \eq{eq1}. The
hierarchical wavefunction for electrons is composed of the fractional
statistics transformation from the pseudo wavefunctions \cite{BW}.
We extract $s$-th level's pseudo-wavefunctional part from
the hierarchical wavefunction on the sphere \cite{Li} as follows:
\beq
  \psi^{(s)}(\zt) = \prod_{i<j = 1}^{N_s}
    d_{ij}(\zt_s)^{\mbox{sgn}(q_{s-2})\theta_{s-1} + p_s},  \label{eq23}
\eeq
with
\beq
d_{ij}(z) = \frac{z_i - z_j}
        {( 1+z_i \zb_i)^{1/2}( 1+z_j \zb_j)^{1/2}},     \label{eq24}
\eeq
where quasiparticle's coordinates are $\zt_s=z_s$ ($\zb_s$) for $s$ odd
(even). $q_n$ and $\theta_n$ correspond to the charge and statistics with
respect to ($n+1$)-th level's quasiparticles
\beq
q_{n+1}=
     - \frac{q_n}{p_{n+1} - \mbox{sgn}(q_n)\theta_n},\hskip 20pt
\theta_{n+1} =
   \frac{\mbox{sgn}(q_n)}{p_{n+1}-\mbox{sgn}(q_n)\theta_n}, \label{eq25}
\eeq
and $q_0 = -1$, $\theta_0 = 0$. $p_1$ is an odd integer and $p_n$ ($\geq2$)
is an even positive integer referring to the index on $n$-th level
pseudo wavefunction such as $m$ on the the Laughlin function in the
above discussion.

According to \eq{eq25}, the $p_n$ for each level determines the statistics
of the subsequent upper levels. Thus, for all fixed $p_n$ ($n<s$), we can
construct a $\Uq$ matrix representation on the $s$-th level
pseudo-wavefunctional bases. Instead of \eq{eq16}, the upper bound of
$p_s$ should be derived from a physical condition for a quantum liquid
state \cite{laugh}. The quantum group generators are similarly given by
\eq{eq17} just replacing $N_e\ra N_s$, $z\ra\zt_s$ with $-\alpha=\beta=1/2$
and the deformation parameter is exactly fixed from the condition
$E^{+}\ket{p_s}=0$ as
\beq
                 q= \exp(2\pi i/p_s).        \label{eq26}
\eeq
This is a similar result as the Laughlin system of $N_e $ electrons.

In this paper, we have discussed various representations of the quantum
group $\Uq$ symmetry only on the pseudo wavefunctional bases separately
from other quasiparticle wavefunctions without a hierarchical structure.
Taking account of a quantum group symmetry on whole hierarchical electron
wavefunction, we could relate the deformation parameter to the continued
fractional filling
\beq
\nu_e ={1\over p_1 + {1\over p_2 + {1\over \dots+{1\over p_s}}}}.
\eeq
In this case, we will need a different representation from this paper's
operators. This should be investigated as a further problem.
\newpage
\noindent
{\em Acknowledgment}

We would like to thank F. Ferrari, W. Zwerger and T. Kobayashi for
useful discussion.


\end{document}